# The Distance-Decay Function of Geographical Gravity Model: Power Law or Exponential Law?

Yanguang Chen

(Department of Geography, College of Urban and Environmental Sciences, Peking University, Beijing 100871, P.R. China. E-mail: chenyg@pku.edu.cn)

**Abstract**: The distance-decay function of the geographical gravity model is originally an inverse power law, which suggests a scaling process in spatial interaction. However, the distance exponent of the model cannot be explained with the ideas from Euclidean geometry. This results in what is called dimension dilemma. In particular, the gravity model based on power law could not be derived from general principles by traditional ways. Consequently, a negative exponential function substituted for the inverse power function to serve for a distance-decay function for the gravity model. However, the exponential-based gravity model goes against the first law of geography. This paper is devoted to solve these kinds of problems by mathematical reasoning and empirical analysis. First, it can be proved that the distance exponent of the gravity model is essentially a fractal dimension. Thus the dimensional dilemma of the power-based gravity model can be resolved using the concepts from fractal geometry. Second, the exponential function indicates locality and localization, which violates the basic principle of spatial interaction. The power function implies action at a distance, which is the necessary condition of geographical gravitation. Third, the gravity model based on power law decay can be derived from the entropy- maximizing principle by introducing a proper postulate. The observational data of China's cities and regions are employed to verify the theoretical inferences, and the results support power-law distance decay. A conclusion can be reached that the preferred form of geographical gravity model is its original form, which is based on an inverse power law rather than a negative exponential law.

*Many of our theories in physical and human geography are being reinterpreted using ideas from fractals and tomorrow they will become as much a part of our education and experience as maps and statistics are today*.

Michael Batty (1992)

# 1 Introduction

The gravity model is one of basic models of spatial analysis in geography and social physics. It provides an empirically effective approach to modeling spatial interaction. The model is originally proposed to describe population migration between two regions (Carey, 1858; Grigg, 1977; Ravenstein, 1885). Afterward, it is employed to measure the force of attraction between any two geographical objects such as cities, firms, and retail stores. Today, the model can be found in many subjects like economics and sociology. A lot of variants of the model came out, and different forms of gravity models have different spheres of application (Erlander, 1980; Haynes and Fotheringham, 1984; Sen and Smith, 1995). All the models remain essentially the same despite apparent changes. The key rests with the law of distance decay of spatial interaction. Despite its practical effect, many basal problems are still pending and remain to be further explored. As is known, the geographical gravity model is initially found by analogy with Newton's law of universal gravitation (Wilson, 2000), thus its distance-decay function is an inverse power function. Nowadays, the inverse power law plays an important role in the studies on complexity. However, for traditional geography, the power law is an obstruction in the way of theoretical construction.

A power function indicates a relation of proportionality between two measures such as length, area, and volume. The geometric measure relations result in a principle of dimension consistency. Two quantities are proportional to one another if and only if they bear the same dimension. The Euclidean dimension must be an integer (0, 1, 2, and 3). Unfortunately, many power laws and power-based models such as the law of allmetric growth and the original gravity model defy this dimension principle. By dimensional analysis, the distance exponent of the gravity model should be an integer or a ratio of one integer to another integer. However, large numbers of empirical results of the distance exponent failed to support the values predicted by the axiom of dimensional homogeneity. The observed values of the exponent are always neither integers nor ratios of two integers, but random values coming between 0 and 3. This cannot be interpreted by the ideas from the traditional mathematics based on Euclidean geometry. Therefore the gravity model once fell in a dimensional dilemma. To solve the dimension problem, theoretical geographers replaced the power function with an exponential function to serve as a distance-decay function (Haggett *et al*, 1977; Haynes, 1975). This model came from the spatial interaction model derived from entropy-maximization principle by Wilson (1970). In terms of the viewpoint of geographers in those years, Wilson's work provided a firmer theoretical basis for many of the spatial interaction models based on loose physical analogy and "raises the gravity model phoenix-like from the ashes" (Gould, 1972, p696; Haggett *et al*, 1977, page 47).

However, the exponential-based gravity model caused a new problem. An inverse power function suggests an action at a distance, it is proper to reflect the nature of spatial interaction; while a negative exponential function suggests locality and localization, it cannot reflect the interaction between two distant places (Chen, 2008; Chen, 2012). In a sense, the function of exponential decay violates the first law of geography, which was presented by Tobler (1970; 2004). On the other hand, the dimensional problem of the distance exponent can be solved by the ideas from fractal theory (Chen, 2009), which was originated by Mandelbrot (1983). Today, many models based on power laws can be extricated from the dilemma of dimension (Batty and Longley, 1994). The novel distance-decay analyses of recent years relate to power function rather than exponential function (Batty, 2008; Chen, 2009; Rybski *et al*, 2013). Partial sporadic conclusions has been published, by so far, there have been no systematic discussion on distance decay effect of the geographical gravity. This paper is devoted to clarifying the mathematical form of the distance-decay function and reinterpreting the distance exponent with fractal dimension concepts. In Section 2, a new theoretical framework is presented to explain the power-based gravity model. This theoretical framework is constructed using the ideas from fractals and entropy maximization. In Section 3, an empirical analysis of Chinese regions and cities is made to support the theoretical inferences on the fractal gravity model. In Section 4, several related questions are clarified or discussed. Finally, the paper is concluded by summarizing the main points of this study.

## 2 Theoretical models

### 2.1 The fractality of the gravity model

The distance exponent of the geographical gravity model can be interpreted with the concepts of scaling and fractal dimension. Suppose that there exist two places, which are numbered by $i$ and $j$. According to first law of geography (Tobler, 1973), the two places are attracted to each other. The interaction between the two places can be measured with the gravity model

$$I_{ij} = G\frac{P_i P_j}{r_{ij}^b}, \tag{1}$$

where $I_{ij}$ denotes the gravity between places $i$ and $j$, which can be represented with the quantity of the flow from one place to the other, $P_i$ and $P_j$ are the "mass", which can be reflected by the population size of places $i$ and $j$, $r_{ij}$ is the distance between $i$ and $j$, $G$ refers to a proportionality coefficient, and $b$, to the distance exponent. Equation (1) is the original form of the gravity model. Suppose there is a point numbered $x$ between places $i$ and $j$. The gravity between $i$ and $x$ and that between $j$ and $x$ can be expressed as

$$I_{ix} = G\frac{P_i P_x}{r_i^b}, \quad I_{jx} = G\frac{P_j P_x}{r_j^b}. \tag{2}$$

where $I_{ix}$ denotes the gravity between $i$ and $x$, $I_{jx}$ is the gravity between $j$ and $x$, $r_i$ refers to the distance from $x$ to $i$, and $r_j$ to the distance from $x$ to $j$. From equations (2) it follows

$$\frac{P_i/r_i^b}{P_j/r_j^b} = \frac{I_{ix}}{I_{jx}}. \tag{3}$$

If $I_{ix}=I_{jx}$ as given, then

$$\frac{P_i}{P_j} = (\frac{r_i}{r_j})^b, \tag{4}$$

which is isomorphic to the law of retail gravitation derived by Reilly (1931). Reilly's model was revised by Converse (1949), who derived a well-known breaking-point formula from equation (3). A breaking point can be defined as a balanced point between two places. The gravity from the breaking point to one place is equal to that from the point to the other place.

A classical problem is how to interpret the distance exponent, $b$. Clearly, equation (3) indicates a spatial scaling relation. The distances $r$ is a length, which denotes a lineal measure with a dimension $d$=1. Let the dimension of the size measure $P$ equal $D$. According to the geometric measure relation, we have

$$(\frac{P_i}{P_j})^{1/D} = (\frac{r_i}{r_j})^{1/d},  \quad (5)$$

where $d$ is the Euclidean dimension of the distance. If $P$ is a 2-dimensional measure, i.e., $D$=2, then we will have $b=D/d$=2; if the dimension of $P$ is $D$=3, then $b=D/d$=3. However, in empirical work, the calculated value of $b$ is neither 2 nor 3, but a random fraction comes between 0 and 3. This phenomenon cannot be interpreted using the concept from Euclidean geometry and gave rise to the dimension dilemma (Chen, 2009; Haggett *et al*, 1977; Haynes, 1975). In order to get over the conundrum, an exponential-based gravity model is proposed as follows

$$I_{ij} = GP_iP_j e^{-r_{ij}/r_0}, \quad (6)$$

where $r_0$ is a scale parameter of spatial interaction. The advantages of the exponential gravity model are as below: first, it is independent of dimension; second, its underlying rationale is clear because it is derivable from the principle of entropy maximization (Wilson, 1970).

A gravity model is actually based on the action at a distance. However, the exponential-based model does not accord with the principle of spatial interaction. The negative exponential function bears a characteristic scale, $r_0$, but it has no long memory. The characteristic scale is associated with the mean of distances. It can be demonstrated that an exponential decay suggests a locality and a process of localization rather than action at a distance (Chen, 2008). Thus a new problem arose from the substitution of the exponential distance-decay function for the power-law distance-decay function. Today, the dimension problem of power-based model can be easily solved by using the ideas from fractals. If the dimension of the size measure $P$ is regarded as fractional, the observed values of $b$ and be explained. For a fractal object, the dimension $D$ can be arbitrary number varying between 0 and 3. Thus the distance exponent $b=D/d=D$ will range from 0 to 3, that is, 0<$b$<3. Another problem is the underlying rationale of the power-based gravity model. In other word, whether or not it can be derived from a general principle. In fact, the gravity model based on the power law can also be inferred with the method of entropy maximization.

## 2.2 Derivation of the power-law based spatial interaction model

The gravity model can be derived from the spatial interaction model, which can be derived from the principle of entropy maximization. Wilson (1970) derived an exponential-based spatial interaction model using the entropy-maximizing method. In fact, the spatial interaction model based on power law can also be derived with the ideas from entropy maximization if we revise one of the postulates given by Wilson (1970), who assumed that the transport cost is linearly proportional to the corresponding distance. Suppose there is a geographical region which can be divided into $n$ zones (sub-regions). The center of gravity of a zone can be represented by its central city. The maximum number of spatial flows between these zones is $n \times n$. Let the total quantity of the flow be $T$, and the flow from zone $i$ to zone $j$ be $T_{ij}$. Based on the assumption of regional partition, the state number of spatial flows is as follows

$$W(T_{ij}) = \begin{pmatrix} & & & T & & \\ T_{11} & T_{12} & \cdots & T_{21} & \cdots & T_{nn} \end{pmatrix} = \frac{T!}{T_{11}!T_{12}!\cdots T_{21}!\cdots T_{nn}!} = \frac{T!}{\prod_i \prod_j T_{ij}!}, \quad (7)$$

where $i$、$j=1,2,\cdots,n$. Based on the distribution of flows, the state entropy can be defined as

$$H = \ln W(T_{ij}) = \ln T! - \ln \prod_i \prod_j T_{ij}! = \ln T! - \sum_i \sum_j \ln T_{ij}!. \quad (8)$$

Suppose that the total outflow quantity from the $i$th origin, zone $i$, is $O_i$, and the total inflow quantity to the $j$th destination, zone $j$, is $D_j$, and the flow quantity from zone $i$ to zone $j$ is $T_{ij}$. We have two constraint equations: $\sum T_{ij}=O_i$, $\sum T_{ij}=D_j$, from which it follows

$$\sum_i \sum_j T_{ij} = \sum_i O_i = \sum_j D_j = T. \quad (9)$$

Then if the number $n$ is a finite value, the maximizing process of the state entropy will result in a uniform distribution of spatial flows, namely, $T_{ij} \equiv const$, where *const* denotes a constant.

However, if the spatial differentiation of a geographical region is taken into consideration, the result will be different. In theory, if $n \to \infty$, then the distance between the centroid of one zone and that of another zone varies from 0 to infinity. The postulate of infinite number of zones is just a mathematical skill, by which a discrete variable can be converted into a continuous variable (Casti, 1996). In virtue of the continuous variables, a physical/geographical problem can be solved by means of calculus. For simplicity, the variable of distance will be replaced with transportation cost. Different distances result in different costs. The long a distance is, the more the transport cost will be. Suppose the unit cost of the flow from zone $i$ to zone $j$ is $c_{ij}$, and the total cost of all the flow in this region is $C$. Equation (9) can be substituted and we have the third constraint equation $\sum\sum c_{ij}T_{ij}=C$. Assuming the evolutive aim of the regional system is entropy maximization of the flow distribution, we can construct a nonlinear programming model, which was propounded by Wilson (1970). The objective function of the optimization problem is

$$\max \quad H = \ln T! - \sum_i \sum_j \ln T_{ij}!, \tag{10}$$

which bears three equality constraints of limited inflow, outflow, and total cost of transportation. The constraint conditions can be expressed as

$$\text{s.t.} \quad \begin{cases} \sum_j T_{ij} = O_i \\ \sum_i T_{ij} = D_j \\ \sum_i \sum_j c_{ij} T_{ij} = C \end{cases} \tag{11}$$

Finding the solution to equation (10) subject to equations (11) yields the relation between the flow quantity and transport cost. Thus spatial interaction models can be elegantly derived from the programming equations. An approach to solving the optimization problem is to utilize the Lagrange multiplier. Based on equations (10) and (11), a Lagrange function can be constructed as

$$L(T_{ij}) = \ln T! - \sum_i \sum_j \ln T_{ij}! + \sum_i x_i (O_i - \sum_j T_{ij}) + \sum_j y_j (D_j - \sum_i T_{ij}) + \beta(T - \sum_i \sum_j c_{ij} T_{ij}). \tag{12}$$

where $x_i$, $y_j$, and $\beta$ refers to the Lagrange multipliers. The well-known Stirling's formula, $T! = (2\pi)^{1/2} T^{T+1/2} e^T$, can be employed to solve equation (12). If $T_{ij}$ is large enough, an approximate relation can be derived from Stirling's formula, i.e., $d\ln(T_{ij}!)/dT_{ij} = \ln(T_{ij})$. Taking the partial derivative of equation (12) with respect to $T_{ij}$ yields

$$\frac{\partial L(T_{ij})}{\partial T_{ij}} = -\ln T_{ij} - x_i - y_j - \beta c_{ij}. \tag{13}$$

The process of mathematical transformation is based on Stirling's approximate relation given above. According to the related knowledge of calculus, the extremum condition is $\partial L(T_{ij})/\partial T_{ij} = 0$. Thus equation (13) can be converted into an exponential function

$$T_{ij} = e^{-x_i} e^{-y_j} e^{-\beta c_{ij}}, \tag{14}$$

This equation leads to the family of Wilson's spatial interaction models, which are based on the law of exponential distance decay (Wilson, 1970; Wilson, 2000).

The deficiency of Wilson's models mainly is reflected in two aspects. First, the negative exponential decay is not consistent with action at a distance (Chen, 2008). Exponential distance decay defies the first law of geography. Second, geographical facts fail to support the linear relationship between traffic costs and distances. As mentioned above, one of Wilson's assumptions is that the transport costs vary directly as the distances. However, this assumption does not accord with reality. As Haggett (2001, page 400) pointed out: "A more realistic curve for distance costs is convex and non-linear, indicating that transport costs increase, but at a decreasing rate, with distance." The most probable curve of distance costs should confirm to a logarithmic function as

$$c_{ij} = c_1 + c_0 \ln r_{ij}. \tag{15}$$

where $c_0$ and $c_1$ are two constants, and $r_{ij}$ is the distance between the centers of gravity of zones $i$ and $j$. Apparently, the growth rate of the transport cost is in inverse proportion to distance, that is, $dc_{ij}/dr_{ij}=c_0/r_{ij}$. This suggests that the transport cost increases over distance, but the rate of growth is decreasing. This accords with what is actually happening in the real world. Substituting equation (15) into equation (14) yields

$$T_{ij} = e^{-x_i} e^{-y_j} (\frac{r_{ij}}{r_0})^{-\beta c_0} = r_0^\sigma e^{-x_i} e^{-y_i} r_{ij}^{-\sigma}, \tag{16}$$

where $\sigma=\beta c_0$ refers to distance exponent, and $r_0=\exp(-c_1/c_0)$ to the characteristic scale of distance, which relates to a mean distance. Without loss of generality, let

$$d_0^{\sigma/2} e^{-x_i} = A_i O_i, \quad d_0^{\sigma/2} e^{-y_i} = B_j D_j. \tag{17}$$

Thus a power-based spatial interaction model is gracefully derived from equations (10) and (11) as below

$$T_{ij} = A_i B_j O_i D_j d_{ij}^{-\sigma}, \tag{18}$$

which was once empirically applied to transport analysis (Taylor, 1977). Differing from Wilson's original model, the power-based model indicates the action at a distance rather than locality. Substituting equation (18) into the constraint equations $\sum T_{ij}=O_i$ and $\sum T_{ij}=D_j$, which is inside equations (11), yields

$$\sum_j A_i B_j O_i D_j d_{ij}^{-\sigma} = O_i, \quad \sum_i A_i B_j O_i D_j d_{ij}^{-\sigma} = D_j. \tag{19}$$

Thus we obtain the following parameters

$$A_i = 1/(\sum_j B_j D_j d_{ij}^{-\sigma}), \quad B_j = 1/(\sum_i A_i O_i d_{ij}^{-\sigma}). \tag{20}$$

which are the scaling factors of spatial interaction defined by Wilson (1970). All the above derivation is same with that presented by Wilson (1970) but a key postulate and the final result. According to the general knowledge of geography, the transport cost is not in linear proportion to distance. The new postulate implies that the growth rate of transport cost with distance is inversely proportional to distance. By improving a simple postulate, we have another spatial interaction model, which can be employed to explain and predict reality in a better way.

**2.3 Derivation of the power-law based gravity model**

A set of power-law based gravity models can be derived from the revised spatial interaction model through allometric relations. Allometric scaling is a ubiquitous phenomenon in both natural and human systems (Chen, 2014a). Bertallanfy (1968) once derived an *aprior* allometry from the principle of general system theory. In fact, allometric relation is a geometric measure relation (Chen, 2010a). Every system is organized according to certain proportions, and each proportion is associated with a geometric measure relation. Thus, the most probable relation between two geographical measures is the allometric relation (Chen and Jiang, 2009). Suppose that the relations between the inflow/outflow and population sizes follow the law of allometric scaling, that is

$$O_i = \eta P_i^u, D_j = \mu P_j^v. \tag{21}$$

where $\eta$ and $\mu$ refers to proportionality coefficients, and $u$ and $v$ to scaling exponents. Substituting equations (21) into equation (18) yields a gravity model such as

$$T_{ij} = K P_i^u P_j^v r_{ij}^{-\sigma}, \tag{22}$$

where the coefficient $K=\eta\mu A_i B_j$. This the simple form of the gravity model. The attraction force is measured with spatial flows. Equation (22) has been empirically verified by observational data (Machay, 1959). A new finding is that there exists a symmetric form of equation (22) in the follow form

$$T_{ji} = K P_i^v P_j^u r_{ij}^{-\sigma}, \tag{23}$$

where has been overlooked for a long time because that geographers takes it for granted that the output $T_{ij}$ equals input $T_{ji}$. Actually, if and only if $u=v=1$, there is a possible symmetric distribution of flows and $T_{ij}=T_{ji}$ (Batty and Karmeshu, 1983). Generally speaking, $T_{ij}\neq T_{ji}$. Equation (22) and Equation (23) form a pair of dual gravity models. Multiplying equation (22) by equation (23) yields

$$T_{ij}T_{ji} = K^2 P_i^{u+v} P_j^{u+v} r_{ij}^{-2\sigma}. \tag{24}$$

From equation (24) it follows equation (1), the power-based gravity model:

$$I_{ij} = (T_{ij}T_{ji})^{1/(u+v)} = K^{2/(u+v)} \frac{P_i P_j}{r_{ij}^{2\sigma/(u+v)}} = G \frac{P_i P_j}{r_{ij}^b}, \tag{25}$$

where the gravity coefficient and distance exponent can be expressed as

$$G = K^{2/(u+v)}, \quad b = \frac{2\sigma}{u+v}. \tag{26}$$

This suggests that equation (1) can be derived from the power-based spatial interaction model by dint of two allometric relations. According to equation (25) and equations (26), the attractive force $I_{ij}$ is a power function of the product of outflow quantity $T_{ij}$ and inflow quantity $T_{ji}$, and the distance exponent $b$ is proportional to the scaling exponent of the spatial interaction model $\sigma$.

Similarly, the exponential-based gravity models can be derived from Wilson's model. The results are as follows

$$T_{ij} = K P_i^u P_j^v e^{-r_{ij}/r_c}, \tag{27}$$

$$T_{ji} = K P_i^v P_j^u e^{-r_{ij}/r_c}, \tag{28}$$

where $r_c$ is a spatial scale parameter, which is associated with the average value of the distances between $n$ places. Equations (27) and (28) combine to yield

$$T_{ij}T_{ji} = K^2 P_i^{u+v} P_j^{u+v} e^{-2r_{ij}/r_c}, \tag{29}$$

from which it follows the exponential-based gravity model, equations (6), that is

$$I_{ij} = (T_{ij}T_{ji})^{1/(u+v)} = K^{2/(u+v)} P_i P_j e^{-2r_{ij}/[(u+v)r_c]} = G P_i P_j e^{-r_{ij}/r_0}. \tag{30}$$

The gravity coefficient and spatial scale parameter can be expressed as

$$G = K^{2/(u+v)}, \quad r_0 = \frac{(u+v)r_c}{2}. \tag{31}$$

It can be proved that the scale parameter of the exponential-based model, $r_0$, equals half average distance. This suggests that the scale parameter can be estimated with the mean of the distances between the $n$ places, $\bar{r}$, which is defined by

$$\bar{r} = \frac{2}{n(n+1)} \sum_{i=1}^{n} \sum_{j=1}^{i} r_{ij} = \frac{1}{n(n+1)} \sum_{i=1}^{n} \sum_{j=1}^{n} r_{ij}. \tag{32}$$

Thus we have $\bar{r} = 2r_0 = (u+r)r_c$, which will be confirmed by the following empirical evidence.

# 3 Empirical analysis

## 3.1 Materials and methods

The systems of urban and regional systems of China are employed to verify the distance-decay function and the main parameter of the gravity model. The study area involves the entire mainland of China. Three measures are adopted, including city size (mass), traffic mileage (distance), and freight volume (the function of gravity). The city size is measured with total urban population in a region, the traffic mileage is measured with the railway distance between the capital cities of two regions, and the freight volume is measured with the quantity of goods carried by railway trains from one region to another region. The data of urban population is from the 5th population census of China in 2000, the matrix of railway distances of cities is extracted from the mileage table of *Atlas of China Transportation*, and the matrix of railway freight volumes comes from 2001 *Year Book of China Transportation and Communications*. The urban population data were processed by the well-known urban geographer, Yixing Zhou, and one of his co-workers. Zhou and Yu (2004a; 2004b) have published the population data of the 666 China's cities based on the 5th census data. It is easy to sum the urban population of all cities in a region to yield a dataset of regional city sizes. There are 31 administrative regions in Mainland China, which comprise 22 provinces, 5 autonomous regions, and 4 municipalities directly under China's Central Government. Two special administrative districts and a special island of China, Hongkong, Macao, and Taiwan, do not belong to Mainland China. The railway network had not reached to the autonomous region of Tibet and Hainan province until 2000. So, only 29 regions are taken into account, i.e., the sample size is $n=29$, and the tables of railway distances and freight volumes make two 29×29 matrixes.

The ordinary least squares (OLS) method is employed to evaluate the parameters of the gravity models. The reason for this is that the OLS method bears an advantage in estimating the regression coefficients of a linear equation. Other methods such as curvefitting method may get more accurate values for the constants $G$ and $K$, but it often cannot give creditable values of the parameters $u$, $v$, $\sigma$, and $r_c$. In geographical analysis, the slope is more important than the intercept in both theoretical studies and empirical analysis. Because of this, the OLS method is more suitable for estimating the parameters of the models based on power laws than other algorithms.

The parameters of the gravity models can be estimated by the log-linear least squares regression.

Taking the logarithms of equations (22), (23), and (24) yields three multivariable linear equations

$$\ln T_{ij} = \ln K + uP_i + v\ln P_j - \sigma \ln r_{ij}, \tag{33}$$

$$\ln T_{ji} = \ln K + vP_i + u\ln P_j - \sigma \ln r_{ij}, \tag{34}$$

$$\ln(T_{ij}T_{ji}) = 2\ln K + (u+v)\ln P_i + (u+v)\ln P_j - 2\sigma \ln r_{ij}. \tag{35}$$

In virtue of these equations, we can estimate the parameters of the power-based models. Similarly, taking logs on both sides of equations (27), (28), and (29) gives

$$\ln T_{ij} = \ln K + uP_i + v\ln P_j - \frac{1}{r_c} r_{ij}, \tag{36}$$

$$\ln T_{ji} = \ln K + vP_i + u\ln P_j - \frac{1}{r_c} r_{ij}, \tag{37}$$

$$\ln(T_{ij}T_{ji}) = 2\ln K + (u+v)\ln P_i + (u+v)\ln P_j - \frac{2}{r_c} r_{ij}. \tag{38}$$

By means of the three equations, we can compute the parameter values of the exponential-based models using the least squares regression.

### 3.2 Results and analysis

The interregional traffic flow of the 29 Chinese regions falls into two types: outflow and inflow. The former is a kind of output flow ($T_{ij}$), and the latter, input flow ($T_{ji}$). Using the city sizes, $P_i$ and $P_j$, and the railway distances, $r_{ij}$, as three independent variables, and $T_{ij}$ as a dependent variable, we can make a multivariate log-linear regression analysis. Fitting the data to equation (33) yields

$$T_{ij} = 0.003964 P_i^{0.4604} P_j^{0.6790} r_{ij}^{-1.1996}, \tag{39}$$

which is the outflow model. The goodness of fit is about $R^2=0.5227$. The distance exponent can be estimated with equations (26), which gives $b \approx 2*1.1996/(0.4604+0.6790) \approx 2.1056$. The symmetrical expression of equation (39) can be obtained by fitting the data to equation (34), and the inflow model is

$$T_{ji} = 0.003964 P_i^{0.6790} P_j^{0.4604} r_{ij}^{-1.1996}, \tag{40}$$

The goodness of fit and the estimated value of the distance exponent based on the second model are the same with those based on the first model. A least squares regression of equation (35) yields

$$T_{ij}T_{ji} = 0.000015714 P_i^{1.1394} P_j^{1.1394} r_{ij}^{-2.3993}, \tag{41}$$

which is the integrated model. The goodness of fit is about $R^2=0.6167$. The distance exponent is $b \approx 2.3993/1.1394 \approx 2.1056$. Obviously, multiplying equation (39) by equation (40) on the same sides produces equation (41). In light of equation (25), equation (41) can be transformed into the power-based gravity model of Chinese fright flow. Extracting the 1.1394th root of equation (41) on both sides yields the normal gravity model as below

$$I_{ij} = 0.00006082 P_i P_j r_{ij}^{-2.1056}. \tag{42}$$

According to equation (42), the gravity coefficient is about 0.00006082, and distance exponent is $b \approx 2.1056$ with a standard error $s_b \approx 0.083$. This is contrast with the result of Rybski *et al* (2013), who once estimated a distance-decay exponent about 2.5 for Paris and its surroundings.

In order to compare the effect of the power-based gravity model with that of the exponential-based model, we should make regression analyses for the gravity model based on exponential distance-decay. Using the least squares calculation to fit the data to equations (36), (37), and (38) yields

$$T_{ij} = 0.000001562 P_i^{0.4684} P_j^{0.6871} e^{-r_{ij}/1480.9268}, \tag{43}$$

$$T_{ji} = 0.000001562 P_i^{0.6871} P_j^{0.4684} e^{-r_{ij}/1480.9268}, \tag{44}$$

$$T_{ij}T_{ji} = 0.000000000002441 P_i^{1.1554} P_j^{1.1554} e^{-r_{ij}/740.4634}, \tag{45}$$

in which 1480.9268 is the estimated value of the parameter $r_c$. The values of the goodness of fit are 0.4687, 0.4687, and 0.5522, respectively. Actually, equations (43) and (44) combine to make equation (45), from which it follows the exponential-based gravity model of Chinese fright flow

$$I_{ij} = 0.00000000008917 P_i P_j e^{-r_{ij}/855.5950}, \tag{46}$$

which gives the estimated value of the spatial scale parameter, $r_0=855.5950$. The average distance can be estimated as $\bar{r}^* = 2r_0 = 1711.1900$. The mean of the observational distances can be computed by using equation (32), which yields $\bar{r} \approx 1780.5885$. Clearly, the $\bar{r}$ value is close to the $2r_0$ value. This suggests that the scale parameter is associated with the mean of distances between different places. The $r_0$ indicates a characteristic length of exponential-based spatial interaction.

From the results of the regression analysis, two conclusions can be drawn as follows. **First, the inflow quantity of a region is not equal to its outflow, but there exists symmetrical** **relationships between gravity models for inflow and those for outflow.** Though the inflow of a zone does equal its outflow, the total inflow equals total outflow of all the zones. Because of this, the inflow gravity models are symmetric to the outflow gravity models. In practice, we need a pair of dual gravity models to describe different flows. **Second, the gravity model based on a power function is better than the model based on an exponential function where the traffic flows of China in 2000 are concerned.** The modeling effect can be evaluated by statistic tests. Both the global tests and local tests show that the power-based gravity model is more suitable for describing the interregional freight flows of china (Tables 1 and 2).

Table 1 Comparison of global effect of the power-based gravity model with that of exponential-based gravity model of China's interregional freight flows (2000)

| Model | Goodness of fit ($R^2$) | Standard error | Degree of freedom | F Statistic | Significance F |
|---|---|---|---|---|---|
| Power-based model | 0.6167 | 1.6333 | 808 | 433.3758 | 9.8017E-168 |
| Exponential-based model | 0.5522 | 1.7655 | 808 | 332.0667 | 1.9172E-140 |

Note: The standard error suggests the prediction effect of a linear regression model.

Table 2 Comparison of local effect of the power-based gravity model with that of exponential-based gravity model of China's interregional freight flows (2000)

| Model | Parameter | | Statistic summary | | | Multi-collinearity statistic | |
|---|---|---|---|---|---|---|---|
| | Item | Coefficient | Standard error | t Statistic | P-value | Tolerance | VIF |
| Power-based model | Intercept | -11.0612 | 1.9939 | -5.5475 | 3.9281E-08 | | |
| | $\ln r$ | -2.3993 | 0.0946 | -25.3725 | 6.4863E-105 | 0.9687 | 1.0323 |
| | $\ln P_1$ | 1.1394 | 0.0758 | 15.0312 | 3.2742E-45 | 0.9823 | 1.0180 |
| | $\ln P_2$ | 1.1394 | 0.0758 | 15.0312 | 3.2742E-45 | 0.9823 | 1.0180 |
| Exponential-based model | Intercept | -26.7385 | 1.9192 | -13.9320 | 1.0679E-39 | | |
| | $\ln r$ | -0.0014 | 0.0001 | -20.8438 | 1.5085E-77 | 0.9655 | 1.0358 |
| | $\ln P_2$ | 1.15549 | 0.0820 | 14.0888 | 1.8061E-40 | 0.9806 | 1.0198 |
| | $\ln P_1$ | 1.15549 | 0.0820 | 14.0888 | 1.8061E-40 | 0.9806 | 1.0198 |

Note: The standard errors are for evaluating the precision of the linear regression coefficients, VIF means "variance inflation factor".

The derivation of the gravity model from the spatial interaction model is based on an assumption that there exist allometric relationships between the size (regional city population) and the flow (traffic flow) of the *n* zones. We can investigate the allometric scaling by fitting equations

(21) to the observational data of China (2000). The first allometry is the power-law relationship between total urban population ($P_i$) and total outflow quantity ($O_i$). The scatterplot shows that the allometric model does not fit the observations very well. The goodness of fit is only $R^2$=0.2743 (Figure 1(a)). However, three aspects of facts should be noted. First, the scattered points are not random distribution. There is a significant trend. Second, among various possible functions depicting the trendline, the power function is the most probable one. Third, the significance level corresponding to the goodness of fit of the allometric model is about 0.0061, which suggests the level of confidence is more than 99.39% (Table 3). The second allometry indicates the scaling relationship between total urban population ($P_j$) and total inflow quantity ($D_j$). The goodness of fit is about $R^2$=0.6127. The power law is the most probable model among various possible relations. The significance of fit is about 0.00000052, which implies the confidence level is greater than 99.99% (Table 3). The effect of the second allometric curve fitting is good in spite of the large deviation. It is bias rather than deviation that we must avoid in a statistical analysis. All in all, the two allometric models are acceptable according to the general criterion of confidence (significance level $\alpha$=0.05).

The difference of the fit quality of the two allometric models reflects the state and property of China's economic and urbanization. The goodness of fit of the second model is significantly greater than that of the first model Figure 1(b)). This suggests that the urban population of a region is mainly correlated with inflow quantity rather than outflow quantity. The inflow reflects the inward attraction of a region, while the outflow reflects regional outward action and influence. China is a developing country. Both urbanization and industrial development are at the primary or intermediate stage (Chen, 2010b). The freight flow comprises principally raw materials such as coal, iron ore, and foodstuff, which come directly from earth surface. In this case, there are many outliers of allometric relation in the scatterplots. For example, Shanxi province is a less developed region in China. Its urbanization level is about 34.91% in 2000. However, Shanxi is the most important area of coal producing. The outflow of Shanxi is not proportional to its urban population due to coal output. Thus, Shanxi forms a typical outlier, a protrudent point, which is displayed at the top left corner of the first subplot of Figure 1(a). The situation of Inner Mongolia (autonomous region) is similar to that of Shanxi. Another typical example is Shanghai (municipality), which represented the most developed area in Mainland China. The urbanization level of 2000 year is about 88.31%. However, the freight flow from Shanghai is relatively small so that the outflow is not very proportional to its urban population. Thus Shanghai makes another kind of outlier. The case of Guangdong province is similar to Shanghai. If we remove the outliers from the dataset, the goodness of fit will go up greatly. In fact, geographical laws are the laws of evolution rather than the laws of existence. The more developed a geographical system is, the more significant a scientific law ruling the system takes on.

**Table 3 The goodness of fit and corresponding significance of the possible models for the relation between regional city population and freight outflow/inflow of China (2000)**

| Relation | Urban population ($P_i$) and Outflow ($O_i$) | | Urban population ($P_j$) and inflow($D_j$) | | Testing |
|---|---|---|---|---|---|
| | Goodness of fit $R^2$ | Significance | Goodness of fit $R^2$ | Significance | $R^2_{0.01,27}$ |
| Linear | 0.0449 | 0.2698 | 0.2753 | 0.0035 | 0.2214 |
| Exponential | 0.0916 | 0.1105 | 0.3323 | 0.0011 | 0.2214 |
| Logarithmic | 0.1290 | 0.0557 | 0.3837 | 0.0003 | 0.2214 |
| Power | 0.2473 | 0.0061 | 0.6127 | 0.0000 | 0.2214 |

**Note**: The fit of freedom of the regression modeling is $n-2=27$. If the level of significance is set to $\alpha=0.05$, the threshold of the coefficient of determination is $R^2_{0.01, 27}=0.2214$. This is to say, if the goodness of fit is greater than 0.2214, then the level of confidence of the scaling exponent will be greater than 99%.

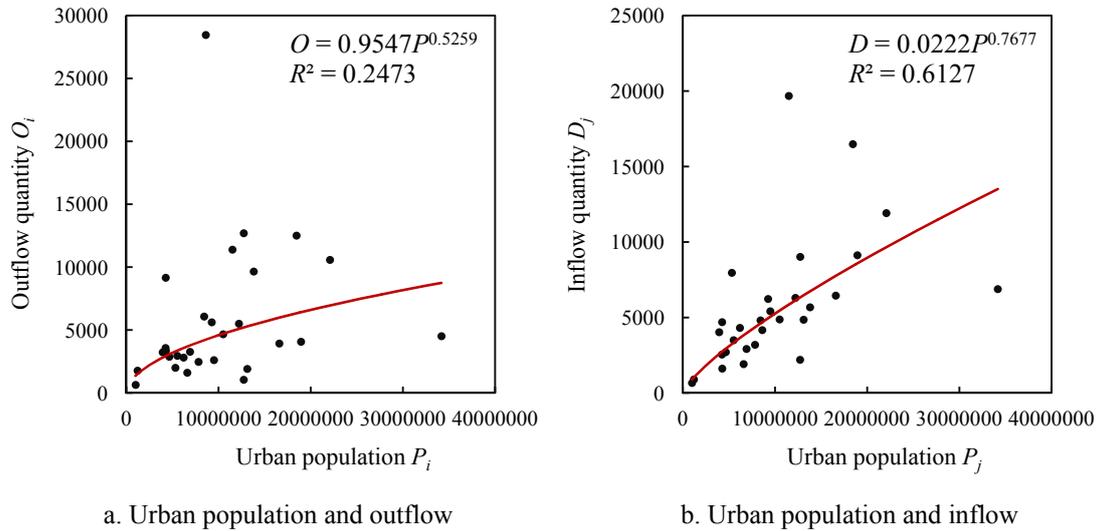

a. Urban population and outflow       b. Urban population and inflow

**Figure 1 The allometric scaling relations between total urban population and total freight volumes (outflow and inflow) of China's 29 regions (2000)**

# 4 Questions and discussion

The gravity model based on power law proved to be associated with fractal dimension and allometric scaling. Assuming that the transport cost is linearly proportional to distance, Wilson (1970) derived the exponential-based spatial interaction model using the entropy-maximizing method. The transport gravity model based on exponential decay can be inferred from Wilson's spatial interaction model. Postulating that the rate of growth of transport cost with distance is in inverse proportion to distance, I derive the power-based spatial interaction from the principle of entropy maximization; then postulating allometric relations between inflow/outflow and population sizes of different places, I derive the power-based gravity model. The dimension problem of the gravity model based on power law can be solved by using the ideas from fractal dimension. However, the locality of the spatial action is still a difficult problem for the exponential-based gravity model. An inevitable trend of development is a regression of the exponential-based gravity model to the gravity model based on power law.

The power-based gravity model relate to the underlying rationale of theoretical geography. Both the first law and the second law of geography are actually based on the original gravity model. The first law of geography of Tobler (1970) reads "everything is related to everything else but near things are more related than distant things"; the second law of geography of Arbia et al (1996) asserts "everything is related to everything else, but things observed at a coarse spatial resolution are more related than things observed at a finer resolution". The first law reflects the inverse proportional relation between the gravity and distance, while the second law reflects the direct proportional relation between gravity and population size product. Both the geographical laws are based on action at a distance. However, the exponential-based distance decay defies the long-distance effect but support the localization. The negative exponential bears a characteristic scale $r_0$. If the distance is greater than the special scale, i.e., $r>r_0$, the gravity of a place following the negative exponential law will attenuate rapidly and then fade out. If the spatial action is dominated by an inverse power law, the gravity will tail off but never vanish. In fact, the distinction between the power-law decay and exponential decay can be revealed by means of autocorrelation function (ACF) and partial autocorrelation function (PACF) analysis. ACF includes direct autocorrelation and indirect autocorrelation, while PACF only denotes direct autocorrelation, no indirect autocorrelation. Both the ACF and PACF of the power-law decay tail off gradually (Figures 2(a) and 2(b)). As for the exponential decay, the ACF tails off (Figures 2(c)), but the PACF cuts off after the first order displacement (Figures 2(d)) (Chen, 2008; Chen, 2012). This suggests that the direct spatial action of a place based on exponential decay is significantly localized and cannot reach distance places. The differences and similarities of the two distance-decay functions have been previously discussed for many times (Batty and Kim, 1992; Chen, 2014b). A simple comparison is drawn between the two types of gravity models (Table 4).

Table 4 Comparison between the exponential-based gravity model and the power-based gravity model

| Items | Exponential-based gravity model | Power- based gravity model |
| --- | --- | --- |
| Distance-decay function | Negative exponential function | Inverse power function |
| Dimension | Euclidean dimension | Fractal dimension |
| Spatial pattern | Simple network | Complex network |
| Spatial process | Localization | Action at a distance |
| Spatial memory | No memory | Long-range memory |
| Symmetry | Scale translational symmetry | Scaling symmetry |
| Basic principle | Entropy maximization | Dual entropy maximization |

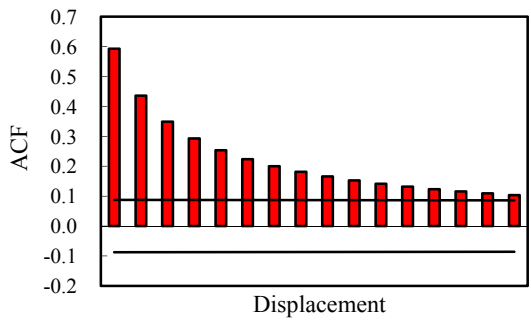

a. ACF of power-law decay

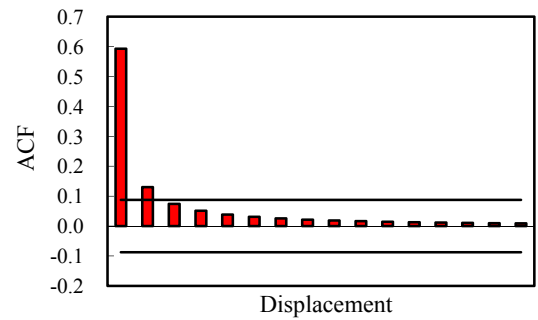

b. PACF of power-law decay

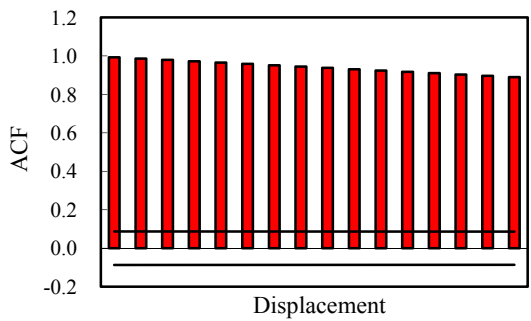

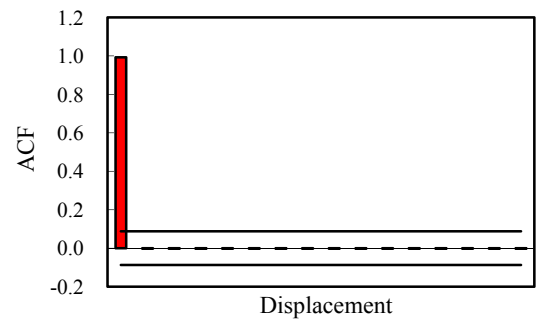

c. ACF of exponential decay          d. PACF of exponential decay

**Figure 2 The histograms of spatial ACF and PACF based on the inverse power function and negative exponential function (by Chen, 2012)**

**Note**: The two horizontal lines or approximate horizontal lines represent the "two-standard-error bands" of ACF or PACF. If the ACF or PACF comes between the two lines, the autocorrelation or partial autocorrelation can be omitted at the significance level $α=0.05$.

One of the defects of this study rests with the assumption of the cost-distance relation. The single logarithmic function on the relation between the distance and transport cost has not been theoretically proved using mathematical reasoning and empirical verified using observational data. The inverse function of a logarithmic function is an exponential function. In principle, the logarithmic relation is also derivable using entropy-maximizing method. However, the process of derivation remains to be researched in future. In fact, how to make a postulate depends on a scientist himself, but the model based on the postulate must be able to explain the observed facts and predict the unknown facts (Komuro, 2001). Compared with the exponential-based model, the power-based model can better explain the human geographical systems. Both urban and regional systems are complex spatial systems (Allen, 1997; Wilson, 2000). It is power laws rather than exponential laws that can be employed to interpret spatial complexity of human geographical systems (Batty, 2005; Batty and Longley, 1994; Frankhauser, 1994). The empirical case is also deficient owing to absence of available data of passenger flow, which results in an incomplete component of traffic flow. Despite this, the calculations are revealing for our understanding the gravity models. In short, the case study is not best, but we cannot find better one because of difficulty of finding observational data for a social scientific study.

## 5 Conclusions

Fractal geometry provides an important mathematical tool for geography and social physics. Twenty years ago, Batty (1992) predicted that many of our theories in physical and human geography could be reinterpreted using ideas from fractals. Actually, the basic geographical models based on power laws such as the allometric growth and gravity models were once abandoned because of dimensional dilemma in theory rather than uselessness in practice. Today, by means of the notions from fractal geometry, we can retrieve these useful power-based models. In a sense, fractal theory raises the gravity model based on the power-law distance decay phoenix-like from the ashes. From the mathematical derivation and empirical analysis, the main conclusions can be drawn as follows.

**First, the gravity model based on power law can be derived from the principle of entropy maximization.** For the spatial variable from 0 to infinity, entropy-maximization results in negative exponential distributions. The exponential-based spatial interaction model and gravity model are derivable using the entropy maximizing methods. A power-law distribution can be deduced from a pair of exponential distributions. This suggests that a power law is based dual entropy-maximizing processes, which are of unity of opposites in self-organized evolution. Replacing the linear distance-cost relation with a logarithmic relation between transport cost and distance, we can derive the power-based spatial interaction model from entropy-maximization hypothesis of regional systems. From the spatial interaction model, we can further derive the gravity model by introducing two allometric realtions. The inverse function of a logarithmic function is an exponential function, which is theoretically derivable from the ideas of entropy maximization.

**Second, the most proper distance-decay function of the gravity model is the inverse power function rather than the negative exponential function.** Both the exponential decay distribution and power-law decay distribution can be derived from general principles, and have firm theoretical basis. The dimension problem of the power-based gravity model can be solved with the ideas from fractals. The distance exponent of the gravity model is a fractal dimension of size measures, and the scaling exponent of the corresponding spatial interaction model can be associated with the fractal dimension of size measures. On the other hand, the locality of the exponential-based gravity model cannot be solved at present. The locality is incompatible with the action at a distance, and thus does not support spatial interaction. What is more, the power-based gravity model can be employed to interpret the complex patterns of geographical systems. However, the exponential –based gravity model indicates simplicity rather than spatial complexity.

**Third, the gravity models are different from spatial interaction models, but they can be connected with one another by allometric relations.** The spatial interaction model can be directly derived from the principle of entropy maximization, while the gravity model can be derived from the spatial interaction model by way of the allometric relations between population

sizes and inflow/outflow quantities. The gravity model can be used to measure the strength of association of one place with another place, while the spatial interaction model can be used to analyze the spatial networks and dynamics of flows or migrations. The gravity model states that the attractive force between two places is proportional to the product of the two places' population sizes and inversely proportional to the distance between the centroids of two places. In contrast, the spatial interaction model states that flow or migration between two places is proportional to the product of the inflow quantity and outflow quantity of the two places and inversely proportional to the intervening distances.

**Fourth, the human force of attraction between two places is not equal to the interregional traffic flow quantity, and we need a pair of dual gravity models to describe inflows and outflows.** General traffic flow, including inflow and outflow, immigration and emigration, results from the gravity of a central place and in turn measures the gravity. A flow quantity is the function of gravity in theory, but in technique, the gravity can be treated as a power function of traffic flow quantity or a function of the product of inflow and outflow so the human attraction forces are measurable in spatial analysis. On the other hand, we need dual gravity models to describe interregional inflow and outflow. The relationship between inflow and outflow is asymmetric, but there is symmetric relationship between the gravity models for inflow and that for outflow. From the dual gravity models it follows the normal gravity models, including the power-based model and the exponential-based one.

## Acknowledgement:

This research was supported financially by the National Natural Science Foundation of China (Grant no. 41171129).